\documentclass[twocolumn]{aastex63}
\usepackage{graphics,amsmath,amsbsy}
\usepackage{lineno}

\newcommand{\lapprox} {\, \lower3pt\hbox{$\sim$}\llap{\raise2pt\hbox{$<$}}\,}
\newcommand{\gapprox} {\, \lower3pt\hbox{$\sim$}\llap{\raise2pt\hbox{$>$}}\,}

\definecolor{mrkred}{RGB}{160,0,0}

\newcommand{\new}[1]{\textcolor{black}{#1}}

\definecolor{mrk}{RGB}{0,0,160}

\shorttitle{Modeling the effects of a lightbridge on oscillations in a solar pore}
\shortauthors{Schiavo et al.}
\begin{document}
\title{Modeling the effects of a lightbridge on properties of MHD waves in solar pores}

\author[0000-0002-5082-1398]{Luiz A. C. A. Schiavo}
\affiliation{Department of Mathematics, Physics and Electrical Engineering, Northumbria University, Newcastle NE1 8ST, UK}

\author[0000-0003-2291-4922]{Mykola Gordovskyy}
\affiliation{Centre for Astrophysics Research, Department of Physics, Astronomy \& Mathematics, University of Hertfordshire, Hatfield AL10 9AB, UK}

\author[0000-0002-7089-5562]{Philippa Browning}
\affiliation{Jodrell Bank Centre for Astrophysics, University of Manchester, Manchester M13 9PL, UK}


\author[0000-0001-5414-0197]{Suzana S. A. Silva}
\affiliation{Plasma Dynamics Group, Department of Automatic Control and Systems Engineering, The University of Sheffield, Mappin Street,
Sheffield, S1 3JD, UK}

\author[0000-0002-9546-2368]{Gary Verth}
\affiliation{Plasma Dynamics Group, School of Mathematics and Statistics, University of Sheffield, Hounsfield Road, Sheffield, S3 7RH, UK}

\author[0000-0002-3066-7653]{Istvan Ballai}
\affiliation{Plasma Dynamics Group, School of Mathematics and Statistics, University of Sheffield, Hounsfield Road, Sheffield, S3 7RH, UK}

\author[0000-0002-6436-9347]{Sergiy Shelyag}
\affiliation{College of Science and Engineering, Flinders University, Tonsley, SA 5042, Australia}

\author[0000-0002-3219-5004]{Sergey N. Ruzheinikov}
\affiliation{Plasma Dynamics Group, Department of Automatic Control and Systems Engineering, The University of Sheffield, Mappin Street,
Sheffield, S1 3JD, UK}

\author[0000-0002-7863-624X]{James A. McLaughlin}
\affiliation{Department of Mathematics, Physics and Electrical Engineering, Northumbria University, Newcastle NE1 8ST, UK}

\author[0000-0002-0893-7346]{Viktor Fedun}
\affiliation{Plasma Dynamics Group, Department of Automatic Control and Systems Engineering, The University of Sheffield, Mappin Street,
Sheffield, S1 3JD, UK}

\begin{abstract}
Solar pores are ideal magnetic structures for wave propagation and transport of energy radially-outwards across the upper layers of the solar atmosphere. We aim to model the excitation and propagation of magnetohydrodynamic waves in a pore with a lightbridge modelled as two interacting magnetic flux tubes separated by a thin, weaker field, layer. We solve the three-dimensional MHD equations numerically and calculate the circulation as a measure of net torsional motion. 
We find that the interaction between flux tubes results in the natural excitation of propagating torsional Alfv\'{e}n waves, but find no torsional waves in the model with a single flux tube. The torsional Alfv\'{e}n waves propagate with wave speeds matching the local Alfv\'en speed where wave amplitude peaks.





\end{abstract}

\section{Introduction}
\label{s-intro}
\color{red}

\color{black}
The structuring of the magnetic field in the solar atmosphere in the form of flux tubes (pores, sunspots, fibrils, prominences, spicules, coronal loops, etc.) provides an ideal environment for mass and energy transfer between different regions permeated by the magnetic field. These structures are also ideal environments for wave propagation and the transport of energy to the upper layers of the atmosphere. The most prominent manifestations of magnetic structures in the solar photosphere are sunspots and pores, locations of the magnetic flux emergence from the solar interior; the differentiation between them resides in the existence of the penumbra region in the case of sunspots. In addition to the difference in their appearances, these two magnetic structures also differ in their size, lifetime and the average intensity of the magnetic field. While in the darkest parts of the umbra, the magnetic field is of the order of 1.7–3.7 kG \citep{Livingstone2002,Solanki2003}, in pores the magnetic field attains values of about 0.6-1.8 kG \citep{Simon1970,Solanki2003}.

Solar pores are intermediate structures between small-scale magnetic flux concentrations in intergranular lanes and fully developed sunspots with penumbrae \citep{Cameron2007}. This makes them an ideal laboratory for studying wave excitation and propagation, including the dissipation of magnetoacoustic wave energy \citep{Millar2020,Grant2015}, and Alfv\'en waves \citep{Morton2011}. These waves traverse upwards through the layers of the lower solar atmosphere along the pore's length which serve as conduits for magnetohydrodynamic (MHD) waves \citep{Morton2011}. Moreover, magnetic pores act as waveguides, transmitting significant wave energy to the upper atmosphere and thereby influencing the dynamics and energetics of the lower solar atmosphere \citep{Keys2018}. Observational evidence has revealed magnetohydrodynamic oscillations within solar pores, discernible in line-of-sight velocities, intensities, and magnetic field strengths \citep{Nelson2021}.

\begin{figure}[htb!]
\centering
\includegraphics[width=0.99\columnwidth]{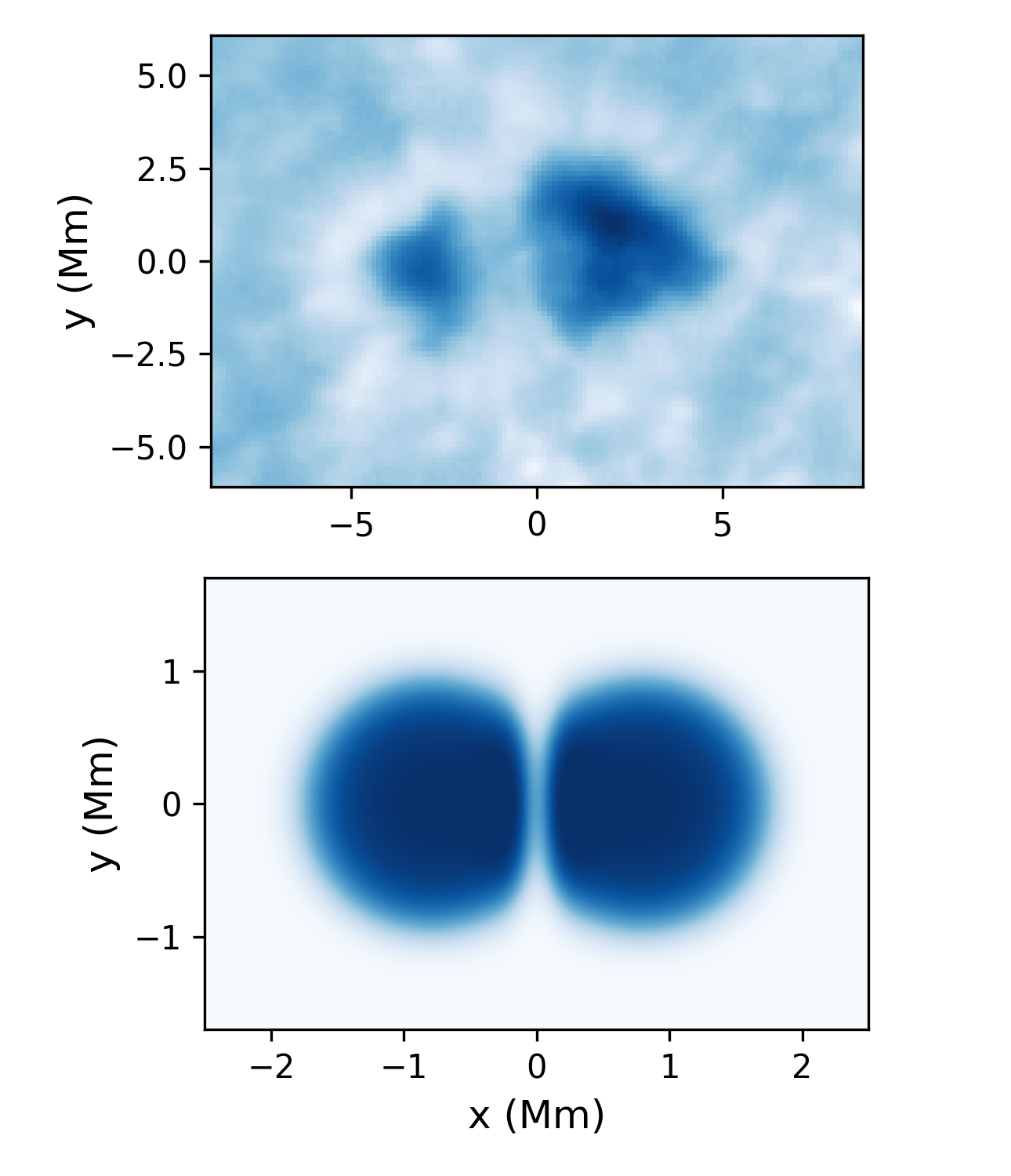}
\caption{Comparison between an observed pore with a lightbridge and the lightbridge numerical model. Upper panel shows continuum intensity for a pore with a lightbridge in active region AR11005 analyzed by \citet{Stangalini2021}. This pore was observed on October 15, 2008 at 16:30~UT at 25.2 N, 10.0 With the Interferometric BIdimensional Spectrometer (IBIS) at the Dunn Solar Telescope (New Mexico, USA). The lower panel presents brightness distributions in our lightbridge model.}
\label{fig:reference-data}
\end{figure}

Some sunspots and pores exhibit lightbridges -- relatively bright elongated structures cutting across sunspot's or pore's umbra (see Fig. \ref{fig:reference-data}, left panel for an example). Observations reveal that lightbridges may have a very versatile magnetic structure, which depends on the overall magnetic structure and evolution of the active region. A lightbridge can be formed as a result of two magnetic elements coming together during the sunspot evolution, leaving a region with a lower vertical magnetic field at the boundary between the two elements, with the force balance in the lightbridge maintained by enhanced (compared to sunspot umbra) gas pressure \cite[see e.g.][and references therein]{boic11, fele16, jine23}. Lightbridges appear to posses a predominantly vertical magnetic field, although the magnetic field strengths in lightbridges appear to be significantly lower compared to the umbrae and, in some cases may be significantly inclined \cite[see e.g.][]{fele16}. In some cases there might be an overlying magnetic field forming magnetic a canopy or a fibril elongated above a lightbridge \cite[e.g.][]{tore15}. The magnetic field in lightbridges is very inhomogeneous, and, in some cases, lightbridges may have locations with extremely strong magnetic field (5--10~kG), i.e. much stronger than in umbrae, although with significantly lower filling factors \cite[$\sim$0.25 compared to $\sim$1 in umbrae, e.g.][]{dure20, loze22}.

Lightbridges in pores have similar characteristics to those in sunspots: their field is mainly vertical, although more inclined than in umbrae, and the force balance is maintained by increased gas pressure \citep{sobe13,kame23}. Hence, in the first approximation, the pores with lightbridges can be considered as two intense magnetic flux tubes joined together.

One notable characteristic of lightbridges is the enhanced power of chromospheric oscillations typically observed in the frequency range of 3-5 mHz \citep{Sobotka2013}. Recently \cite{Stangalini2021} detected torsional Alfv\'en waves in a Fe I spectral line in a pore under the presence of a lightbridge. They further emphasised the importance of torsional Alfv\'en waves for chromospheric and coronal heating estimating their energy flux. \cite{Stangalini2021} also performed a numerical simulation of a flux tube driven by a kink driver to generate torsional waves. However, they considered a single tube model for their numerical simulations, which is not consistent with the observed magnetic configuration of a solar pore with lightbridge.

In this work, a photospheric pore with a lightbridge is modeled numerically as two closely-adjacent magnetic tubes, which represent magnetic elements with a thin layer of weaker magnetic field and higher pressure separating them (see Fig.~\ref{fig:reference-data}). For the analysis of the wave propagation, and in order to focus on the essential physics of waves driven by the presence of the lightbridge configuration compared with a single pore, we neglect gravitational stratification and the large-scale velocity field, therefore our initial configuration is independent of height \new{for simplicity}. Although the employed model is relatively simple, the configuration adequately reflects the main properties of a sunspot or a photospheric pore with a lightbridge. Individual oscillating flux tubes driven by kink drivers have already been studied by \cite{Terradas2008,Pascoe2010,Antolin2014}.

An attempt to model a realistic situation in the solar atmosphere was to consider magnetic flux tubes as being built up from a multiple of cylindrical structures that show collective motion. Earlier studies by, e.g. \cite{Terradas2008b,Ofman2009,Robertson2011,Soler2015,Magyar2016,Shi2024} showed that in such systems the interaction between individual structures modifies the spatial structures and morphology of waves. Here, for the first time, we are simulating two closely-adjacent magnetic elements forming a lightbridge under kink motion.

\section{Numerical simulations}
\subsection{Main equations}
\label{s-setup}

We perform three-dimensional ideal magnetohydrodynamic (MHD) simulations of oscillations in a model representing a pore and a pore with the lightbridge using the Lare3D code where MHD equations are solved in the Lagrangian form employing a Lagrangian-Eulerian remap procedure \citep{Arber2001}. The equations solved by Lare3D are presented in dimensionless form as
\begin{eqnarray}
\frac{D\rho}{D t} &=& - \rho \nabla \cdot \mathbf{v} , \\
\rho\frac{D\mathbf{v}}{D t} &=& (\nabla \times \mathbf{B} ) \times \mathbf{B} - \nabla p + \mathbf{f}_{visc} , \\
\frac{D\mathbf{B}}{D t} &=& (\mathbf{B}\cdot\nabla) \mathbf{v} - \mathbf{B}(\nabla \cdot \mathbf{v}), \\
\frac{D\epsilon}{D t} &=& - \frac{p}{\rho} \nabla \cdot \mathbf{v}, \\
p &=& \rho \epsilon(\gamma -1),
\label{eq:mhd}
\end{eqnarray}
where $\mathbf{v}$ denotes the velocity vector, $\mathbf{B}$ represents the magnetic field, $\rho$ indicates plasma density, $p$ corresponds to gas pressure, $\epsilon$ is the specific internal energy, and $\gamma$ is the ratio of specific heats, here set to 5/3. To address numerical instabilities and manage steep gradients, such as shocks, a numerical viscosity vector, $\mathbf{f}_{visc}$, is introduced \citep{Arber2001,Caramana1998}.

The model assumes a fully ionized plasma, and the governing equations are normalized by a length-scale $L_0$, magnetic field strength $B_0$, and density $\rho_0$. These three constants are further utilized to establish normalization for velocity ($v_0 = B_0/\sqrt{\left.\mu_0 \rho\right.}$), pressure ($p_0=B_0^2/\mu_0$), time ($t_0=L_0/v_0$), specific internal energy scales ($\epsilon_0=v_0^2$), and temperature ($T_0=\epsilon_0\overline{m}/k_B$), where $\mu_0$ is the vacuum permeability, $k_B$ is the Boltzmann constant and $\overline{m}$ is the average ion mass, defined as 1.2 times the proton mass. While the simulation results can be scaled with any suitable reference scales, they are defined here to align with typical values of the photosphere, specifically $L_0 =$~1~Mm, $B_0 =$~0.17~T, and $\rho_0~=~1.67\times10^{-4}$~kg~m$^{-3}$. Therefore, the normalization velocity and temperature are $v_0=11.7$~km~s$^{-1}$ and $T_0=20,000$~K, respectively, and the scale time is $t_0=85$~s.

In our simulations, the normalized computational domain box size is $8\times6\times16$ in the $x$, $y$, and $z$ directions, respectively.
 The simulation box is extended in the $z$-direction to 16 to minimize the effect of wave reflections from the upper boundary. However, our analysis only considers the domain between $z=0$ and $z=8$. Each considered model covers a time period that was over 150 Alfv\'en times.

\subsection{Initial and boundary condition}

We consider two models: a ``reference'' configuration with a single, cylindrical magnetic flux tube, and one with two partially merged flux tubes with magnetic field depression between them. The latter configuration is used to simulate a photospheric pore with a lightbridge. 

\begin{figure*}[htb!]
\centering
\includegraphics[width=0.99\textwidth]{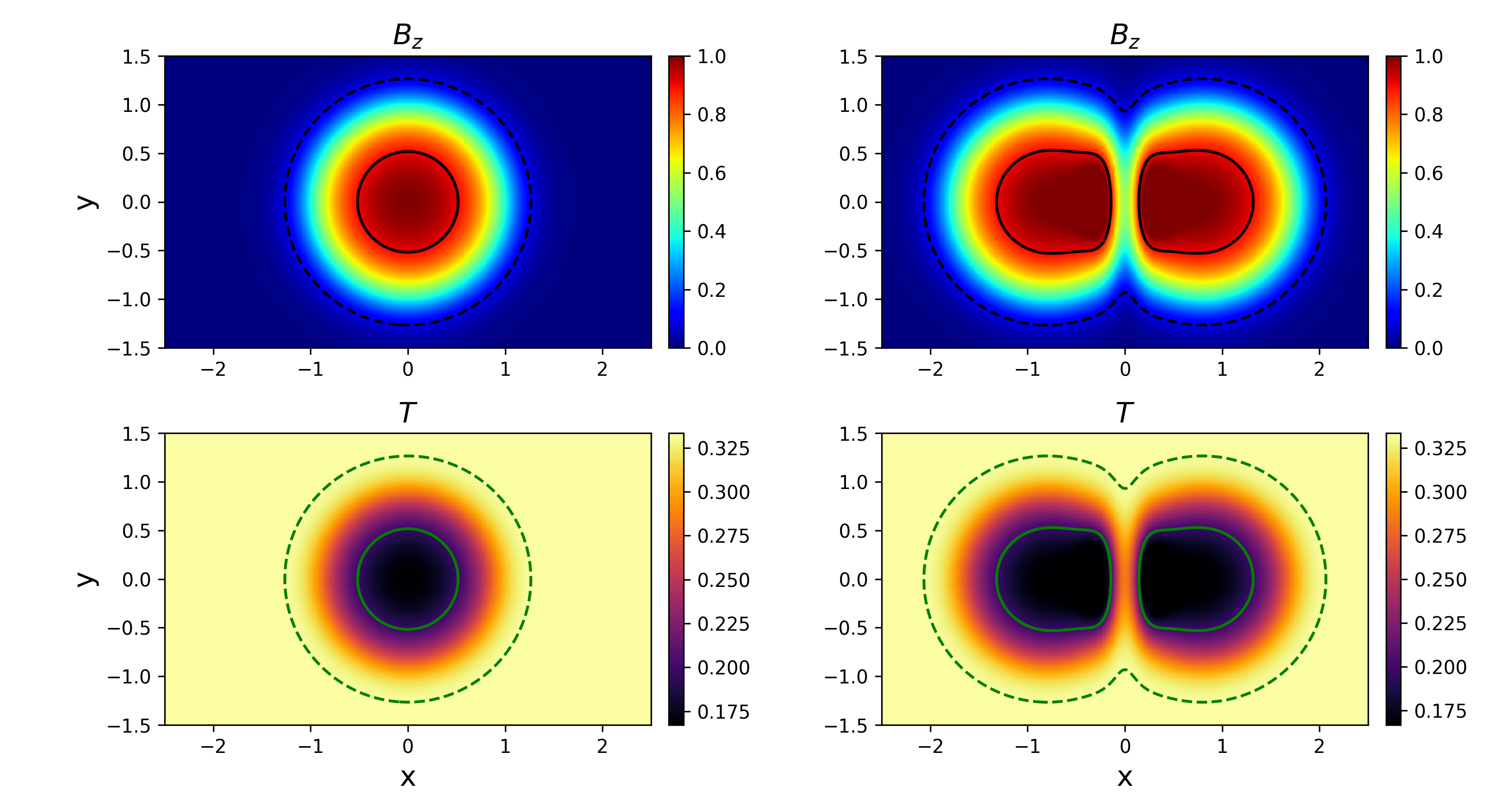}
\caption{
The initial conditions for the magnetic field, $B_z$ and the continuum intensity for the single flux tube model (left panels) and the lightbridge model (right panels). Dashed lines and solid lines represent iso-contours of $B_z=$~0.1 and $B_z=$~0.9. The upper panels show contours of $z$-component of the magnetic field. The lower panels show contours of temperature, where the normalizing scales are $B_0$=~0.17~T and $T_0=$~20000~K.}
\label{fig:initial-condition}
\end{figure*}

The initial magnetic field and temperature for these configurations are shown in Figure \ref{fig:initial-condition}.

The inhomogeneous initial magnetic field in the single flux tube model is given as a two-dimensional Gaussian function centred on the origin
\begin{equation}
B(x,y) = B_0 \exp\left(-\frac{x^2+y^2}{R_0^2}\right),
\end{equation}
while in the case of the lightbridge model, the magnetic field is given as
\begin{eqnarray}
B(x,y) = &B_0& \left[\exp\left(-\frac{(x-x_s)^2+y^2}{R_0^2}\right) \right. \nonumber \\ 
&+& \exp\left(-\frac{(x+x_s)^2+y^2}{R_0^2}\right) \nonumber  \\
&-& \left. \exp\left(-\frac{32 x^2+y^2}{R_0^2}\right)\right].
\end{eqnarray}

The latter term in this equation is used to create a magnetic field depression between the two partially merged fluxtubes. This region with the magnetic field strength lower and the temperature higher than inside the fluxtubes, represents the lightbridge (see the right panels in Figure~\ref{fig:initial-condition}. Here $R_0=1.0 L_0$ is the flux tube radius. The parameter $x_s$ which determines the relative positions of the flux tubes in the lightbridge model is taken to be $0.8 L_0$. The initial density is assumed  to be constant $\rho(x,y,z,t=0) = \rho_0$, while the pressure distribution is obtained assuming the magnetohydrostatic equilibrium condition:
\begin{eqnarray}
\rho(x,y,z,t=0) &=& \rho_0,\\
p(x,y,z,t=0) &=& p_0 - \frac{B^2(x,y,z,t=0)}{2\mu_0},\\
\epsilon(x,y,z,t=0)&=& \frac {p(x,y,z,t=0)}{\rho_0(\gamma-1)}. 
\end{eqnarray}
The ambient pressure value is set to $p_0= B_0^2/ \mu_0$. Hence, the gas pressure and temperature inside the fluxtubes is $\sim 2$ times lower than outside.
 
The left panels show the magnetic configuration with a single flux tube, while the right panels correspond to the magnetic configuration with two partially overlapping flux tubes, with the overlap region with magnetic field depression representing the lightbridge. The solid lines in Figure \ref{fig:initial-condition} correspond to iso-contours of $B_z=0.9$, effectively, representing the cores of the fluxtubes. 

Periodic boundary conditions are used at the side boundaries of the model domains ($x=-4$, $x=4$, $y=-3$ and $y=3$). At the upper boundary of the numerical domain, we use Neumann boundary conditions, setting a zero gradient for every variable. At the lower boundary ($z=0$), the same zero gradient conditions are imposed on all variables apart from $v_y$. The $v_y$ component of the velocity, which is used as a driver, is given as
\begin{equation}
v_y(x,y,z=0,t) = A \sin\left(\frac{2\pi}{\lambda} t\right) g(x,y),
\end{equation}
where $A$ denotes the driver amplitude that was set at a value of 0.05 to guarantee $v_y$ perturbations stay in the linear regime and the function $g(x,y)$ is set as
\begin{equation}
g(x,y) = 0.5 \left[1 + \tanh\left(\frac{R_d-\sqrt{\frac{x^2}{4}+y^2}}{0.2}\right)\right].
\end{equation}
This function is a constant equal to $1$ in the middle of the lower boundary, in a region with the radius $R_d=1.5$, and reduces to zero near the side boundaries. Therefore, the bases of the fluxtubes in both models oscillate in $y$-direction as one whole, while there driver's velocity is zero at the edges of the lower boundary.

The quantity $\lambda$ represents the oscillation period or the period of the driver (chosen as 10$t_0$, equals to 850~s, corresponding to the frequency of 1.2~mHz).

\section{Results}
\subsection{Flow evolution}

\begin{figure*}[htb!]
\centering
\includegraphics[width=0.87\textwidth]{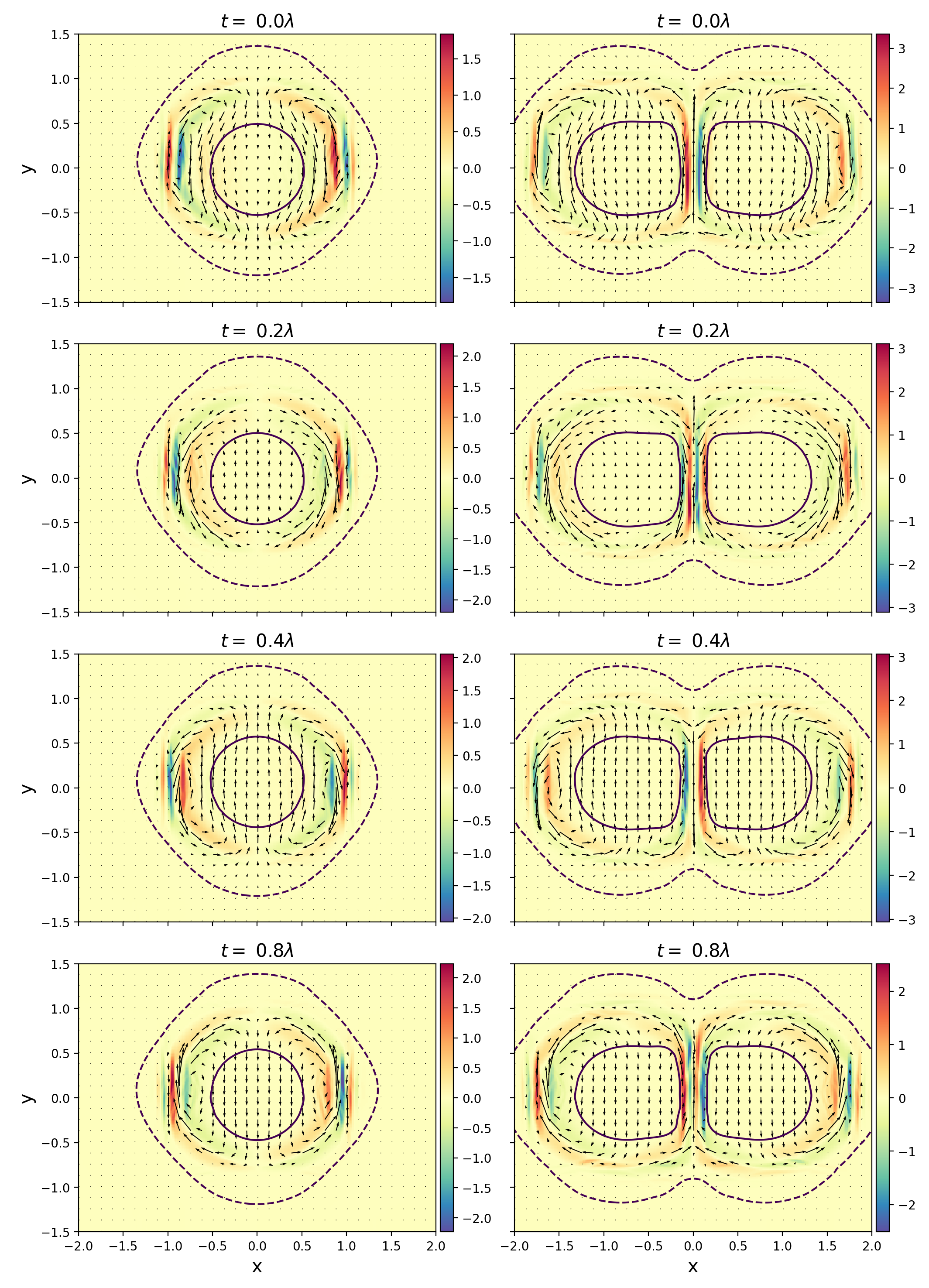}
\caption{Contours of $z$-component of vorticity measured at $z=8$. The arrows represent the velocity vector at different simulation times. The left panel represents the single tube simulation while the right panel the lightbridge model. $\lambda$ represents the period of perturbation.} 
\label{fig:velocity-vectors}
\end{figure*}

Before analyzing the behavior of the flux tubes, we will examine the evolution of vorticity and magnetic field lines over an oscillation cycle. Figure \ref{fig:velocity-vectors} illustrates the $z$ component of the vorticity field, $\omega_z=(\nabla\times {\bf v})_z$, and velocity vectors at four successive simulation time steps within an oscillation period for the cases of a single flux tube (left column) and lightbridge simulations (right column). As before, the dashed and continuous lines represent the isocontours of the vertical component of the magnetic field, $B_z$, set at 0.1 and 0.9, respectively. 

The vorticity is prominently observed in a boundary layer outside the core of the flux tube, extending to its external boundary. This vorticity generation seems to be associated with the kinking motion of the flux tube core. In the case of the single flux tube, two symmetric vortices are observed at $x = \pm 0.8$ and $y = 0$, while in the lightbridge case, there are four vortices along the $x$ axis. Despite this difference, the vorticity profiles remain symmetric, albeit with a higher amplitude in the case of a sunspot with a lightbridge. Periodic changes in the rotational patterns between clockwise and counterclockwise directions characterize the small swirling motions observed at $y = 0$ along the $x$ axis. These periodic changes in swirl orientation create a wave pattern in the vorticity contour, which is visible at the edges of the flux tubes. This vorticity oscillation generates a periodic torsional motion that can be associated with a torsional Alfv\'en wave. The field lines in the flux tube core exhibit a kink mode behavior characterized by a near-zero $B_x$ component, indicative of minimal vortical motion.

\begin{figure}[htb!]
	\centering
	\includegraphics[width=\columnwidth]{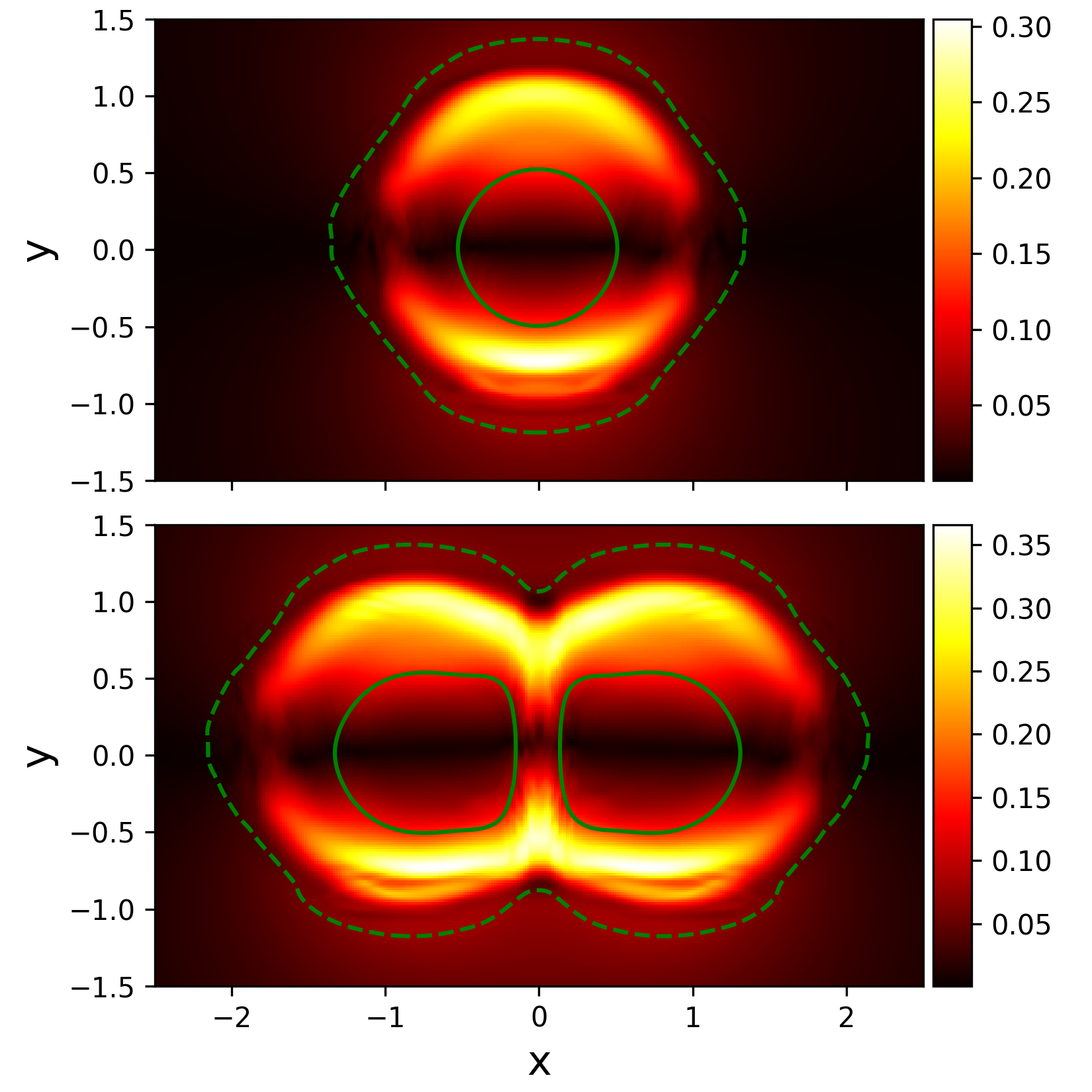}
	\caption{Contours of PSD of $v_z$ measured at $z=4$ \new{ at the diver frequency, upper panel displays results for a single flux tube and the lower panel for the lightbridge model. The lightbridge case presents a higher power between the pore core and the boundary. }} 
	\label{fig:psdvz}
\end{figure}

Figure \ref{fig:psdvz} illustrates the outcomes of a power spectral density (PSD) analysis conducted for $v_z$ at $z = 4$. The contour plots depict the PSD results for each grid point at the driver frequency, $f_D=1/\lambda$. \new{As usual, the maximum of PSD corresponds to regions where the power of waves takes its maximum value.} To avoid the initial transient, these PSDs were computed using time series data collected from $t = 20$ to $t = 100$.

\new{The PSD contour maps for a single flux tube show a strong signal for $v_z$ between the flux tube core and its boundary, indicating wave propagation. The lightbridge case displays a similar pattern, but we can see that the interaction between flux tubes creates an intense signal in the lightbridge.}

The lightbridge case shows, a prominent signal between the boundary and the pore core. However, this signal diminishes towards the middle of the lightbridge, a finding consistent with those reported by \cite{Stangalini2019} who noted a lack of noticeable amplitude in the velocity line-of-sight power spectral density at the lightbridge in observational data for frequencies of 6 mHz and 8 mHz,  and no signal from circular polarization. 

\subsection{Net circulation}
\begin{figure*}[htb!]
\begin{center}
\includegraphics[trim= 85 0 0 0,clip, width=0.99\textwidth]{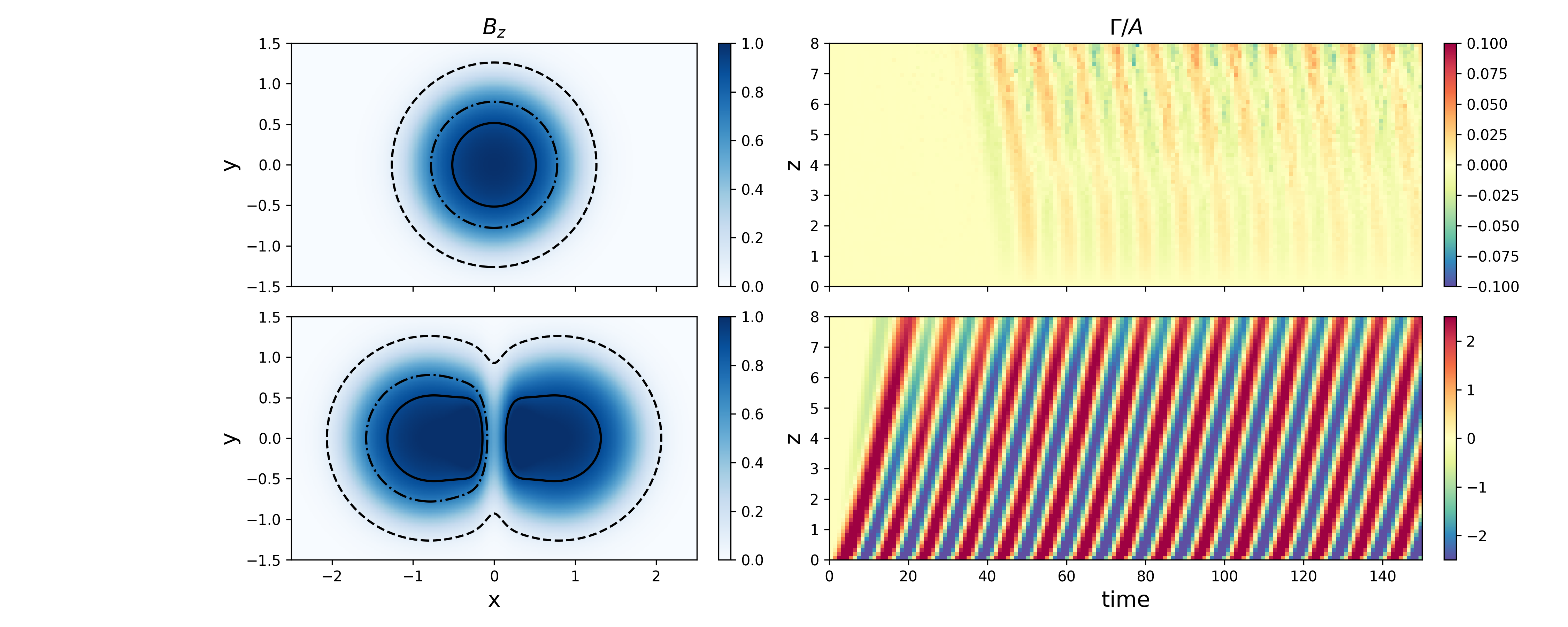}
\caption{The upper row shows results for the single flux tube simulation and the bottom row for the lightbridge simulation. Left panels: contour plots of $B_z$. The dashed-dot line delimits the \new{path of} integration employed to compute circulation, the dashed line represents $B_z=$ 0.1 and the solid line $B_z=$ 0.9. Right panels: time-distance diagram of the circulation normalized by the driver amplitude, $A$. }
\label{fig:circulation-diagram}
\end{center}
\end{figure*}

In order to detect the possible existence of azimuthal motions associated with torsional Alfv\'{e}n waves, we compute the circulation, $\Gamma$, for each simulation as a measure of net torsional motion. The circulation, well-known in fluid dynamics, is defined as the line integral of a velocity vector field around a closed curve, 
\begin{equation}
\Gamma = \oint_{C} \mathbf{v}\cdot d\mathbf{l} =
\iint_S \nabla \times \mathbf{v}\cdot d\mathbf{S} = 
\iint_S \pmb{\omega} \cdot  d\mathbf{S} ,
\end{equation}
where $C$ is the closed curve and $S$ the surface defined by the closed curve $C$, and $\pmb{\omega}=\nabla \times \mathbf{v}$ is the vorticity vector. The circulation integrated over $x-y$ planes for each simulation case is shown in Figure \ref{fig:circulation-diagram} at different time steps. The integrating surface, $S$, is chosen to be an isocontour of $B_z=0.7$. The area was chosen to be large enough to capture the vorticity oscillation, while small enough to capture a single lobe of the flux tube. The left column features contour plots of $B_z$, with the dashed-dot line indicating the integration area set to correspond to $B_z=0.7$, and the dashed and continuous lines represent the flux tube boundary and its core. The right column presents the normalized circulation by the driver amplitude, $A$. The upper row corresponds to the single flux tube simulation, while the lower row represents the lightbridge simulation.

As indicated by the upper right panel, the net circulation is negligible in the single flux tube, as expected due to the anti-symmetry of the $z$ vorticity contours shown in Fig.\ \ref{fig:velocity-vectors}. Between $t = 0$ and $t = 40$, the net circulation remains very close to zero. This is expected from the azimuthal symmetry of this configuration, as there is no preferred direction for net torsional circulation. However, after this period, some low-amplitude waves are observed being reflected from the upper boundary towards the bottom of the domain. Fortunately, these waves do not compromise the analysis since their amplitude is ten times smaller than the driver amplitude. By contrast, in the lightbridge case, the net circulation for a single tube is (in each lobe of the lightbridge) no longer zero. In this case, the normalized circulation exhibits an amplitude three times larger than the driver perturbation and propagates as a wave towards the upper boundary at $z=8$. This suggests an upward-propagating torsional Alfv\'{e}n mode with a single frequency and constant velocity. 
This arises from the interaction between the two lobes in the lightbridge simulation leads to an asymmetry in each lobe's vorticity profile, resulting in a net circulation. Equivalently, the azimuthal symmetry in each lobe is broken due to the distortion from the lightbridge, allowing net torsional circulation to develop. \new{The waves reflected in the lightbridge case are not visible on the space-time diagram because they have a small amplitude. The influence of reflected waves can be disregarded in this analysis, as their amplitude is more than ten times smaller than that of the upward propagating wave.}

\begin{figure*}[htb!]
\centering
\includegraphics[width=0.99\textwidth]{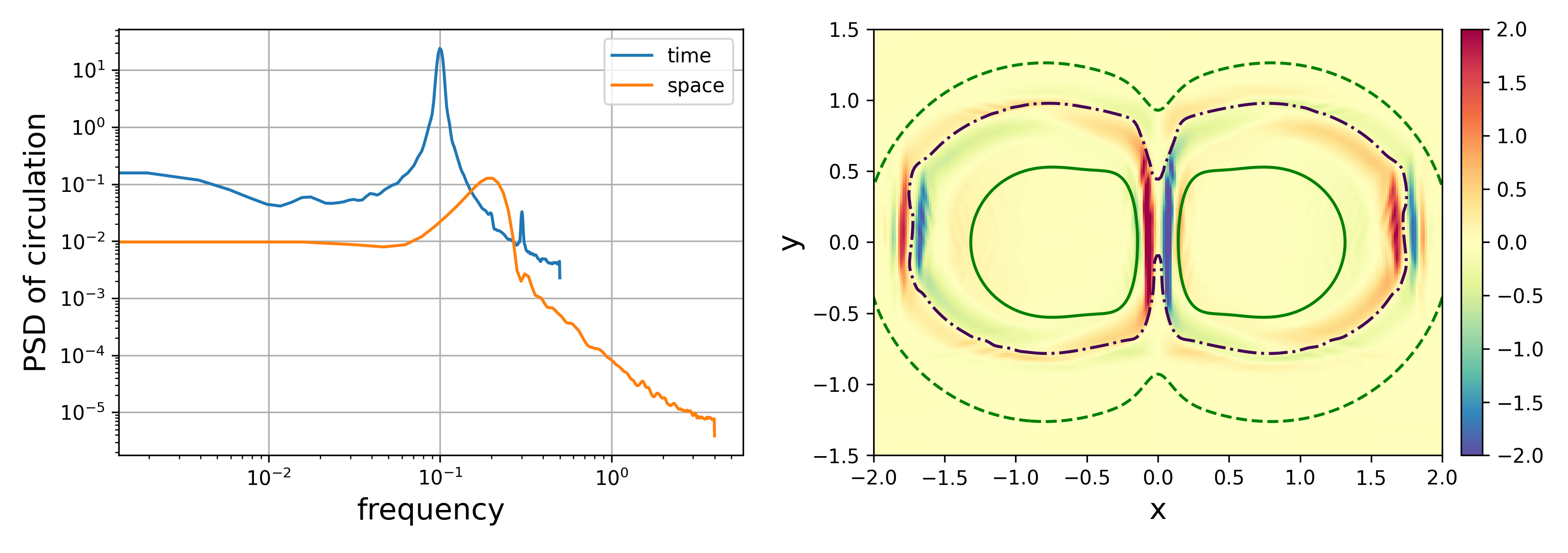}
\caption{Left panel: PSD applied to circulation in space and time. Right panel: plot of $\omega_z$ for $z = 8$ at $t=50$. The black dash-dotted line indicates the region where wave speed is equal to Alfv\'en speed. Green dashed line and solid lines represent isocontours of $B_z=$ 0.1 and $B_z=$ 0.9, respectively.}
\label{fig:psd}
\end{figure*}

Figure \ref{fig:psd} displays the results of the power spectral density (PSD) computed for the spatial-time diagram presented in Figure \ref{fig:circulation-diagram}, specifically for the lightbridge case. The PSD computation involves spatial analysis in the $z$ direction between $t=40$ to $140$, followed by averaging to determine the dominant wavenumber. We selected this particular time interval to keep the analysis free from the influence of initial transients. Temporal PSD is then calculated between $z=3$ to $8$ to mitigate upper and lower boundary effects; the result was also averaged to smooth the signal. The main frequency and the primary wavenumber were identified and used to compute a phase speed of 0.531 $v_0$, corresponding to 11.6~km~s$^{-1}$. The right panel of Figure \ref{fig:psd} shows the values of $\omega_z$ for $z = 8$ at $t=50$. The black dash-dotted lines indicate locations where the propagation speed equals the local value of the  Alfv\'en speed. Notably, the phase speed matches the Alfv\'en speed within the outer boundary layer of the flux tube, and more or less in the location where the vorticity peaks in magnitude, so the wave amplitude is strongest. This suggests a torsional Alfv\'en wave localized within this layer. 

\subsection{MHD modes}
In order to analyze the nature of perturbations in the system, we employ the wave decomposition method described in detail by \cite{mumford2015}. In the context of linear perturbations in a uniform homogeneous magnetized plasma, the ideal MHD equations have three independent eigenmodes corresponding to the fast and slow magneto-acoustic waves and the Alfv\'en waves. These modes exhibit distinct properties, with characteristics dependent on the plasma conditions in which the waves propagate. Decomposing perturbations into these modes is a non-trivial problem.

\begin{figure*}[htb!]
\centering
\includegraphics[width=0.87\textwidth]{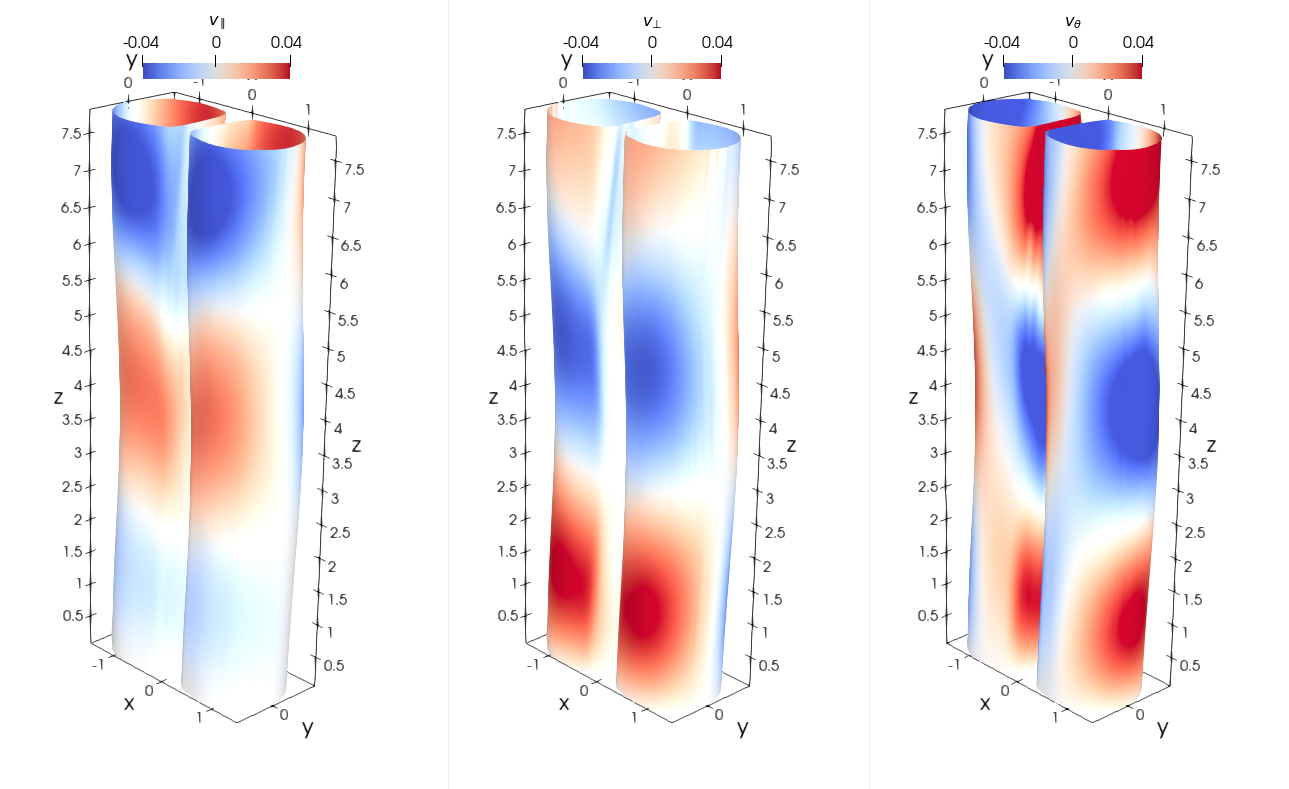}
\caption{Contours of velocity field decomposed into parallel, $v_\parallel$, perpendicular, $v_\bot$, and azimuthal, $v_\theta$ components. }
\label{fig:velocity-decomposition}
\end{figure*}

Identifying the three modes of oscillation in a 3D geometry becomes feasible in the presence of flux tubes. The fast, slow, and Alfv\'en modes can be associated with velocity perturbations perpendicular to the magnetic field and the flux tube, perturbations parallel to the flux tube and its surface, and an azimuthal vector perpendicular to the magnetic field and parallel to the surface, respectively. \cite{mumford2015} proposed the following decomposition for the energy flux
\begin{eqnarray}
F_\parallel &&= \rho v^2_\parallel  c_S \ , \label{eq:energy1} \\
F_\bot &&= \rho v^2_\bot v_A \ , \label{eq:energy2} \\
F_\theta &&= \rho v^2_\theta v_A \ ,
\label{eq:energy3}
\end{eqnarray}
where $F_\parallel$, $F_\bot$, and $F_\theta$ are the parallel, perpendicular and azimuthal components of the energy flux, $v_\parallel$, $v_\bot$, and $v_\theta$ are the parallel, perpendicular and azimuthal velocity components, and $c_S$, $v_A$ are the sound and Alfv\'en speeds. The azimuthal component computed by $\mathbf{n}_\theta = \mathbf{n}_\bot  \times \mathbf{n}_\parallel$, where $\mathbf{n}_\parallel$ is the \new{unit} vector parallel to magnetic field and $\mathbf{n}_\bot$ perpendicular to the surface constructed numerically within a computational domain as an isosurface where $|\mathbf{B}|=0.8$. Given the assumption that perturbations are small, we consider the isosurface to be parallel to the magnetic field lines. This 3D surface allows us to obtain the vectors $\mathbf{n}_\theta$ and $\mathbf{n}_\bot$. In Figure \ref{fig:velocity-decomposition}, we present the velocity field decomposed into parallel ($v_\parallel$), perpendicular ($v_\bot$), and azimuthal ($v_\theta$) components. Notably, the perpendicular velocity component exhibits a larger amplitude near the driver and decays along the $z$ direction, while the azimuthal and parallel components demonstrate similar amplitudes along $z$ direction.

\begin{figure*}[htb!]
\centering
\includegraphics[width=0.87\textwidth]{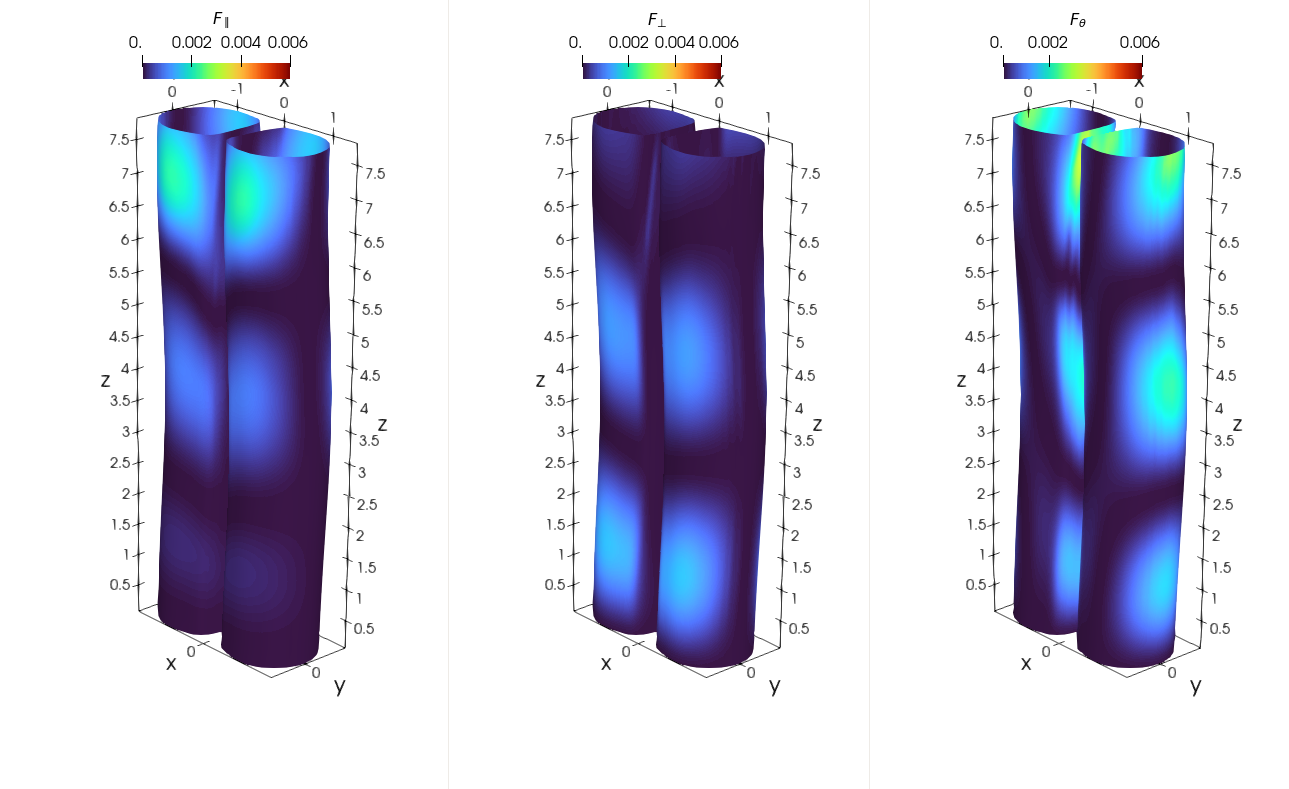}
\caption{Contours of energy flux field decomposed into parallel, $F_\parallel$, perpendicular, $F_\bot$, and azimuthal, $F_\theta$, components. }
\label{fig:energyflux}
\end{figure*}

In Figure \ref{fig:energyflux}, we show the energy flux decomposition based on \new{Eqs.\ (\ref{eq:energy1})-(\ref{eq:energy3}).} The kink driver induces a peak in the perpendicular energy flux component, $F_\bot$, near the bottom of the domain, which propagates upwards and attenuates rapidly. The component $F_\parallel$ has its minimum at $z=0$ and increases with $z$. In the absence of any dissipative effect, it is likely that $F_\bot$ is being converted into $F_\parallel$ as it propagates to higher altitudes. The azimuthal energy flux, $F_\theta$, is generated at the lower boundary by the kink driver, with an apparent intensification around $z=7$. The driver contributes to both perpendicular and azimuthal perturbations, which means torsional waves and kink waves; such behavior has also been observed by \cite{mumford2015} and \cite{Stangalini2021}.
\new{In our model, plasma $\beta$ is less than one, which means that the $F_{\parallel}$, $F_{\perp}$, and $F_{\theta}$ are associated with the dominant eigenfunctions representing a slow magnetoacoustic, fast magnetoacoustic and Alfv\'en waves, respectively \citep{Jess2015}. Therefore, since $\theta$ is the azimuthal component, $F_{\theta}$ will be associated with the energy flux of torsional Alfv\'en wave.}

\label{s-model}

\section{Conclusions}
Our investigation focused on a pore with a lightbridge, which we modeled as two closely-adjacent magnetic flux tubes separated by a thin layer of weaker field. To simplify our analysis and in order to focus on generic properties of wave propagation and generation, we ignored gravitational stratification and large-scale velocity fields, which allowed us to generate a height-independent initial configuration. Despite this simplification, our model effectively captures the primary characteristics of a sunspot or a photospheric pore with a lightbridge, both driven by a kink driver at their bases.

An examination of vorticity and field line evolution revealed that no vorticity is generated within the flux tube core,  where the axial field is 90\% or more than its peak value ($B_z>0.9$), and the field lines exhibit kink body-wave motions. Vorticity emerges between the pore core and its boundary, with the lightbridge scenario demonstrating intensified vorticity near the lightbridge. Vorticity profiles no longer exhibit symmetry within each lobe, in contrast to what is observed in a single flux tube.

In the case of a single flux tube, the net circulation remains minimal due to a cancellation between vorticity profiles of opposite sign on either side of the tube, resulting from symmetry across the $y$ axis. However, the presence of two closely adjacent flux tubes disrupts this balance, leading to a net circulation of significant magnitude in each lobe of the lightbridge. This disruption causes torsional motion on each side due to their interaction, with torsional waves occurring in the boundary layer between the edges of the flux tubes and their cores. Notably, we found that the speed of these waves matches the local Alfvén speed in regions where wave amplitude peaks.

Our analysis of the MHD mode decomposition elucidates that the kink driver engenders both torsional and kink waves. Moreover, the perpendicular energy flux to the flux tube surface decreases with height while the azimuthal component increases.

Importantly, we demonstrated that a single flux tube driven by a kink driver does not exhibit net torsional motion, a result that contradicts the numerical modeling reported by \cite{Stangalini2021}. Our model establishes a new and crucial understanding: net torsional motion arises from the interaction between the two lobes of the pore and that including this structuring in the background model is essential for interpreting waves observed in pores and sunspots with lightbridges. 
\label{s-summary}

\section*{Acknowledgments}
This work used the DiRAC Data Intensive service (CSD3) at the University of Cambridge, managed by the University of Cambridge University Information Services on behalf of the STFC DiRAC HPC Facility. L.S. and P.B. are grateful to the  Science and Technology Facilities Council (STFC) for support from grant ST/T00035X/1, L.S. and J.M. are funded by the STFC ST/X001008/1. M.G. is funded by STFC grant ST/Y001141/1. V.F., G.V. and SSA are grateful to the STFC grants ST/V000977/1 and ST/Y001532/1. V.F., G.V. and I.B. acknowledge the support provided by the Royal Society International Exchanges grants with Greece (IES/R1/221095), Australia (IES/R3/213012) and India (IES/R1/211123). This research has also received financial support from the ISEE, International Joint Research Program (Nagoya University, Japan). The authors thank Dr Marco Stangalini for the useful discussions during our work on this paper and the data provided.
\bibliographystyle{aasjournal}
\bibliography{svb23}

\clearpage

\end{document}